\documentclass[aps,prl,shownopacs,twocolumn]{revtex4}

\usepackage{graphicx}
\usepackage{hyperref}
\usepackage{amsmath}
\usepackage{epstopdf}
\usepackage{bbm}
\newcounter{firstbib}

\begin{document}

\newcommand{\ket}[1]{$\left|#1\right\rangle$}

\title{Waveguide-Coupled Single Collective Excitation of Atomic Arrays}

\author{Neil V. Corzo \footnotemark[2]\footnotetext{\footnotemark[2]Present address: Centro de Investigaci\'on y Estudios Avanzados del Instituto Polit\'ecnico Nacional - Unidad Quer\'etaro, 76230 Quer\'etaro, M\'exico}}
\affiliation{Laboratoire Kastler Brossel, Sorbonne Universit\'e, CNRS, ENS-Universit\'e PSL, Coll\`ege de France, 4 Place
Jussieu, 75005 Paris, France}
\author{J\'er\'emy Raskop}
\affiliation{Laboratoire Kastler Brossel, Sorbonne Universit\'e, CNRS, ENS-Universit\'e PSL, Coll\`ege de France, 4 Place
Jussieu, 75005 Paris, France}
\author{Aveek Chandra}
\affiliation{Laboratoire Kastler Brossel, Sorbonne Universit\'e, CNRS, ENS-Universit\'e PSL, Coll\`ege de France, 4 Place
Jussieu, 75005 Paris, France}
\author{Alexandra~S.~Sheremet}
\affiliation{Laboratoire Kastler Brossel, Sorbonne Universit\'e, CNRS, ENS-Universit\'e PSL, Coll\`ege de France, 4 Place
Jussieu, 75005 Paris, France}
\author{Baptiste Gouraud\footnotemark[3]\footnotetext{\footnotemark[3]Present address: Department of Physics, University of Basel, Klingelbergstrasse 82, CH-4056 Basel, Switzerland}}
\affiliation{Laboratoire Kastler Brossel, Sorbonne Universit\'e, CNRS, ENS-Universit\'e PSL, Coll\`ege de France, 4 Place
Jussieu, 75005 Paris, France}
\author{Julien Laurat \footnotemark[1]\footnotetext{\footnotemark[1]julien.laurat@sorbonne-universite.fr}}
\affiliation{Laboratoire Kastler Brossel, Sorbonne Universit\'e, CNRS, ENS-Universit\'e PSL, Coll\`ege de France, 4 Place
Jussieu, 75005 Paris, France}

\maketitle

\textbf{Considerable efforts have been recently devoted to combining ultracold atoms and nanophotonic devices \cite{nanofiber1,nanofiber2,Lukin2013,phc} to obtain not only better scalability and figures of merit than in free-space implementations, but also new paradigms for atom-photon interactions \cite{RMP}. Dielectric waveguides offer a promising platform for such integration because they enable tight transverse confinement of the propagating light, strong photon-atom coupling in single-pass configurations and potentially long-range atom-atom interactions mediated by the guided photons. However, the preparation of non-classical quantum states in such atom-waveguide interfaces has not yet been realized. Here, by using arrays of individual caesium atoms trapped along an optical nanofibre \cite{Sile,Solano17bis}, we observe a single collective atomic excitation \cite{Duan2001,Sangouard2011} coupled to a nanoscale waveguide. The stored collective entangled state can be efficiently read out with an external laser pulse, leading to on-demand emission of a single photon into the guided mode. We characterize the emitted single photon via the suppression of the two-photon component and confirm the single character of the atomic excitation, which can be retrieved with an efficiency of about 25\%. Our results demonstrate a capability that is essential for the emerging field of waveguide quantum electrodynamics, with applications to quantum networking, quantum nonlinear optics and quantum many-body physics \cite{Kimble,Chang2014}.\\}

The integration of optical emitters and nanophotonic structures has been a major goal in optical physics and quantum information science. Over recent years, micro- and nanoscale lithographed cavities have enabled the confinement of light into smaller volumes and have led to a variety of realizations for single-atom cavity quantum electrodynamics (QED). These remarkable advances include the implementation of efficient solid-state single-photon sources \cite{Sennelart} and of quantum gate functionalities \cite{Tiecke2015}. In parallel to these cavity-based settings, the coupling of emitters to nanoscopic waveguides has also raised a large interest, with diverse platforms based on quantum dots \cite{Lodahl2015}, molecules \cite{Turschmann} or cold atoms \cite{Chang2014}. This waveguide-QED approach, which is also pursued with superconducting qubits and microwave transmission lines \cite{vanLoo2013}, enables the strong interaction of a single or multiple atoms with a guided mode, in a single-pass lossless configuration. 

Besides the promise of scalable integrated quantum nodes and their combination into complex quantum networks for studying emergent phenomena \cite{Schleier}, waveguide-QED systems based on atomic ensembles have spurred an intense theoretical effort to investigate new avenues for light-matter and matter-matter interactions mediated by the guided photons \cite{RMP}. Recent proposals include the engineering of many-body dark states arising from possible chiral coupling to the waveguide \cite{Pichler2015}, or the generation of entangled atomic states that can be mapped to complex multiphoton states with unprecedented fidelity scaling \cite{Tuleda2017}. It has also been suggested that collective effects in trapped arrays enable the exponential improvement of photon storage fidelities using selective radiance \cite{Asenjo2017}. Moreover, optical dispersion engineering of the waveguide and the resulting tunable long-range interaction between the atoms would give rise to new opportunities for quantum non-linear optics and  simulation, including the generation of photon molecules \cite{Douglas2016} and the realization of a large class of spin Hamiltonians and synthetic quantum matter \cite{Douglas2015,Choi2017}. 

On the experimental front, cold atoms have been trapped in the evanescent field of nanoscale dielectric waveguides, such as optical nanofibres \cite{nanofiber1, nanofiber2} and photonic-crystal slow-mode waveguides \cite{phc}. These pionneering works have led to the first demonstrations of all-fibred optical memories \cite{Gouraud2015,Sayrin2015} and the observation of collective effects, such as super-radiance \cite{Goban2015,Solano17} and Bragg reflection \cite{Corzo2016,Polzik2016}. To explore further this nascent field and emerging platforms, the capability to prepare non-classical states of the arrayed atoms is a major challenge that has yet to be realized. 

Here we demonstrate the heralded creation of a waveguide-coupled single collective excitation of arrays of individual atoms. Previously, single collective excitations have been generated in free-space trapped atoms or in doped crystals, and they constitute a key resource for quantum networking tasks and quantum non-linear optics \cite{Kimble,Sangouard2011,Chang2014}. However, preparing such excitation in a chain of individual atoms --that is, an atomic register-- and coupling such a non-classical state preferentially to a waveguide are two capabilities that remain elusive. In our experiment, we demonstrate these abilities and we characterize the collective entangled state by the subsequent on-demand emission of a guided single photon.

\begin{figure}[t!]
\includegraphics[width=0.9\columnwidth]{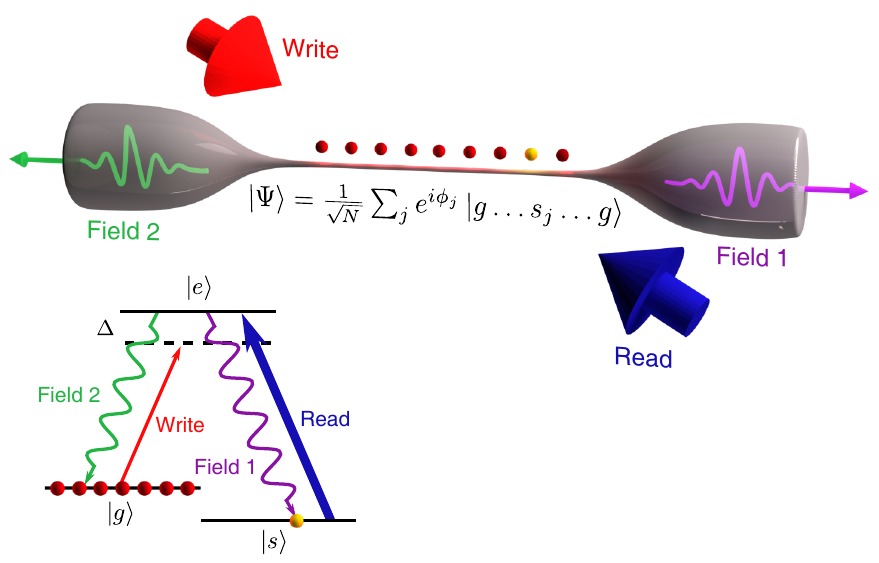}
\vspace{-0.15 cm}
\caption{\textbf{Waveguide-coupled collective excitation in an atomic register.} Via an external write pulse on the $|g\rangle \rightarrow |e\rangle$ transition, a collective entangled state with one `spin-flip' excitation is created into a chain of individual cold atoms trapped in the evanescent field of a nanoscale waveguide. This process is heralded by the detection of a photon in the guided Field-1 mode. At a later time, a read pulse on the $|s\rangle \rightarrow |e\rangle$ transition converts the excitation into a single photon in the guided Field-2 mode. This scheme offers a fibre-addressable single collective excitation $|\Psi\rangle$ in an atomic register with on-demand readout.}
\label{fig1}
\end{figure}

\begin{figure*}[t!]
\includegraphics[width=1.8\columnwidth]{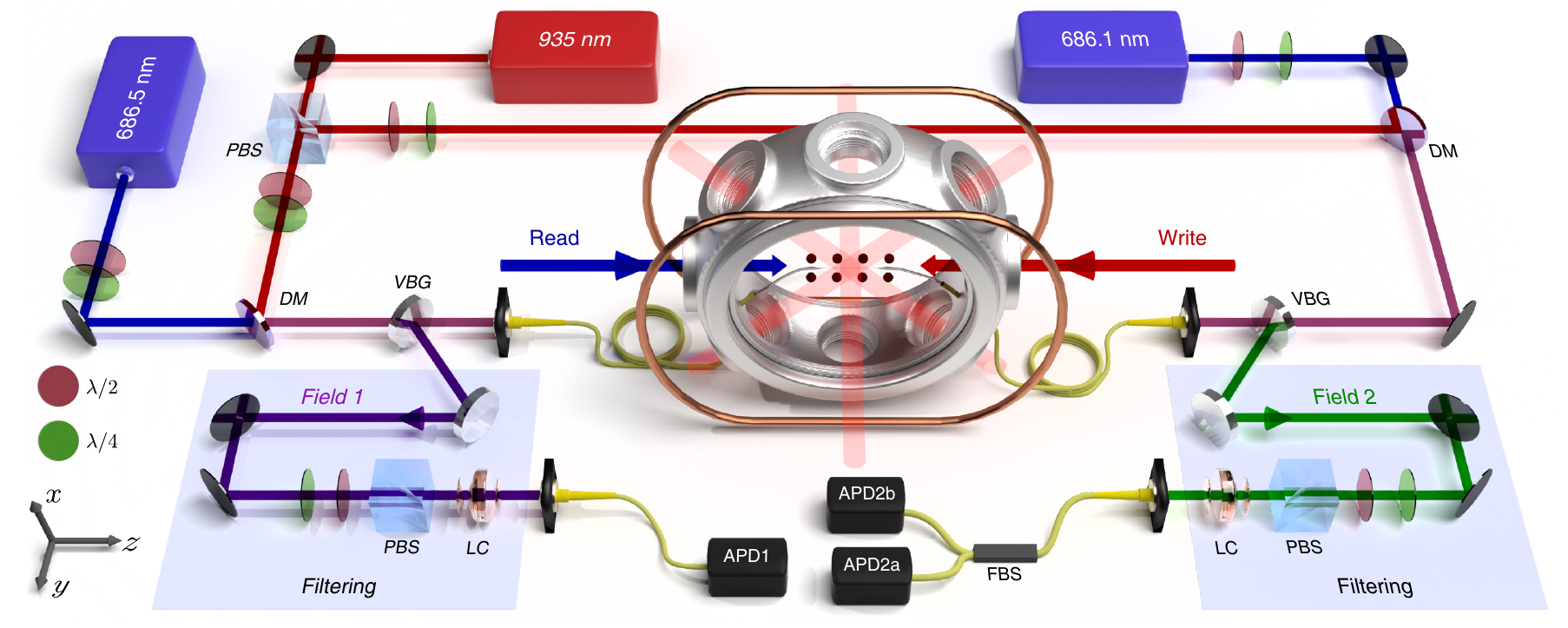}
\caption{\textbf{Experimental setup.} Two arrays of about 1000 cold cesium atoms each are trapped in the evanescent field of a 400-nm-diameter optical nanofibre. The state-insensitive compensated optical lattice is loaded from an elongated magneto-optical trap and realized by two pairs of guided counter-propagating beams at the cesium magic wavelengths. One red-detuned pair provides the attractive potential whereas one blue-detuned pair (with a relative detuning between the two beams) gives a repulsive contribution, overall realizing individual trapping sites 200 nm from the surface. The levels $\{|g\rangle, |s\rangle, |e\rangle\}$ correspond to the Cs hyperfine levels $\{| 6S_{1/2}, F=4\rangle,| 6S_{1/2}, F=3\rangle,| 6P_{3/2}, F'=4\rangle\}$ and the atoms are initially prepared in the $|g\rangle$ ground state. The heralded entangled atomic state is studied by the characterization of photon pairs, i.e., heralding Field-1 and readout Field-2 photons emitted into the guided modes, after frequency filtering via cascaded volume Bragg gratings (VBG), lens cavities (LC) and polarizing beamsplitters (PBS). Photons are detected with single-photon counting modules (APD1, APD2a and APD2b). DM stands for dichroic mirror and FBS for fibre beamsplitter.}
\label{fig2}
\end{figure*}

Our building block is illustrated in Fig. \ref{fig1}. The atomic ensemble consists of a chain of $\Lambda$-type individual atoms trapped around a one-dimensional nanoscale waveguide and addressable via the evanescent tail of the guided mode. After initializing the atoms in the $|g\rangle$ ground state, an external weak write pulse detuned from resonance illuminates the ensemble and induces spontaneous Raman scattered fields. The detection of a single photon in the Field-1 guided mode heralds the creation of a long-lived collective entangled state with one excitation shared among the whole atomic chain. To retrieve the written single excitation, an external read pulse is sent to the atomic ensemble after a programmable delay. The read pulse, which is resonant to the  $|s\rangle \rightarrow |e\rangle$ transition, deterministically maps the excitation into a waveguide-coupled Field-2 single photon. Collective enhancement should enable this state transfer to be efficient, as demonstrated in free-space atomic ensembles following the seminal Duan-Lukin-Cirac-Zoller proposal \cite{Duan2001,Kuzmich03,Laurat07,Sangouard2011}.

The experimental set-up is detailed in Fig. \ref{fig2}. We use a silica single-mode optical fibre and create a 1-cm long tapered region of 400-nm diameter by the standard heat-and-pull technique \cite{Sile,Solano17bis}. The nanofibre is then suspended inside an ultrahigh vacuum chamber and connected to the outside via teflon feedthroughs. The overall transmission is above 98\%. 

In this setting, we trap arrays of laser-cooled cesium atoms in the evanescent field. The all-fibred dipole trap consists of two pairs of guided counterpropagating beams, one attractive red detuned and another repulsive blue detuned, operating at the specific cesium magic wavelengths \cite{nanofiber2}. In this state-insensitive compensated trap, the differential scalar and vector shifts are strongly suppressed. The polarization of the guided beams is aligned by measuring the polarization properties of the Rayleigh scattering from surface imperfections. The four dipole beams are quasi-linearly polarized along the $x$ axis, as shown in Fig. \ref{fig2}. Owing to the evanescent structure of the guided mode, the dipole trap generates two parallel chains of potential minima along the nanofibre waist, one on the top and one on the bottom, with at most one atom per site.

We start the experiment by loading the dipole trap, which is constantly on (See Methods). For this purpose, a cigar-shaped magneto-optical trap is overlapped with the nanofibre axis and the cold atoms are loaded during a 4.5-ms-long optical molasses phase. In all of the following measurements, the optical depth (OD) of the trapped atomic ensemble is maintained at about 20 for a probe on the $|g\rangle  \rightarrow |e\rangle$ transition and polarized along the $x$ axis. This value corresponds to a total number of $2000~\pm~200$ atoms, as estimated via a saturation measurement. The single-atom coupling ratio $\Gamma_{1D}/\Gamma_0$ amounts thereby to about $10^{-2}$, where  $\Gamma_{1D}$ is the radiative decay rate into the guided mode and $\Gamma_0$ into free space. The trap lifetime is measured to be 25~ ms. Additional details about the dipole trap, including its loading and characterization, are presented in Methods.

To create a single collective atomic excitation, we send a weak write pulse propagating in free space. The 50-ns-long write pulse is detuned by  $\Delta=-10~ \rm{MHz}$ from the $|g\rangle \rightarrow |e\rangle $ transition, and horizontally polarized along the $y$ axis. The angle between the write beam and the nanofibre is minimized to a value $\alpha=5~^{\circ}$ and the waist is equal to $1~\rm{mm}$ to overlap the atomic chains uniformly. At this stage, we herald a collective excitation in the ensemble by detecting a single photon in the Field-1 mode. We select the Field-1 mode to be the guided mode of the nanofibre, and quasi-linearly polarized along the $x$ axis. 

To study the stored excitation, a 10-mW external read pulse, resonant with the $|s\rangle  \rightarrow |e\rangle$ transition, is sent to the ensemble after a programmable delay. The read beam is mode-matched to the write beam but travels in the opposite direction, and it is linearly polarized along the $x$ axis, i.e., orthogonal to the write polarization. The read pulse maps the collective excitation into a Field-2 photon that escapes the ensemble, but in the opposite direction of Field 1. The Field-2 mode is also the guided mode but quasi-linearly polarized along the $y$ axis.

To separate the Field-1 and Field-2 photons from the bright dipole lights and the residual leakage of the write and read auxiliary beams guided by the nanofibre, the photons are each directed to a filtering system, composed of two cascaded volume Bragg gratings (VBG), a polarizing beamsplitter and a lens cavity (See Methods). In each filtering system, the total isolation from the dipole beams is 160 dB and the total transmission, including fibre coupling, is 40 \%.  After the filtering, the Field-1 and Field-2 photons are detected by single-photon avalanche photodiodes.
 
\begin{figure}[t!]
\includegraphics[width=0.93\columnwidth]{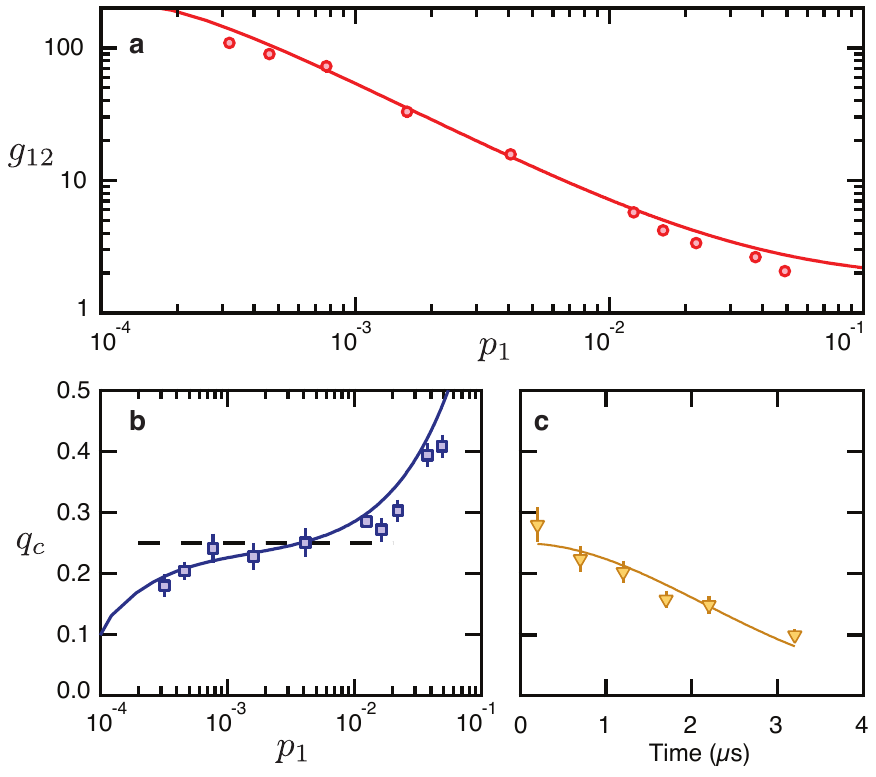}
\caption{\textbf{Characterization of the collective excitation}. \textbf{a}, Normalized cross-correlation function $g_{12}$ between Field 1 and 2, and \textbf{b}, conditional retrieval efficiency $q_{c}$ as a function of the probability $p_{1}$ to detect a heralding photon in Field 1. The solid lines correspond to the theoretical model that assumes the output state to be a two-mode squeezed state with additional coherent and incoherent backgrounds. \textbf{c}, Retrieval efficiency as a function of the storage time. The solid line provides a fit given by $\exp[-(t/\tau)^2]$, with $\tau=3\pm0.3 ~\mu$s. The OD of the atomic ensemble was set to $20$ for the $|g\rangle \rightarrow |e\rangle$ transition, with a probe quasi-linearly polarized along the $x$ axis. The error bars correspond to the propagated Poissonian error of the photon counting probabilities.}
\label{fig3}
\end{figure}

We now turn to the experimental characterization. We first study the non-classical correlations between the generated photon pairs, Field 1 and Field 2, via the normalized cross-correlation function $g_{12}=p_{12}/(p_{1}\cdot p_{2})$, where $p_{12}$ is the joint probability to detect a pair of photons and $p_{i}$ the probability of detecting a photon on field $i$ \cite{Laurat07}. The dependence of $g_{12}$ with $p_1$, i.e., the excitation probability, is presented on Fig. \ref{fig3}a. We observe that the value of $g_{12}$ increases when the excitation probability is reduced, as expected. In addition, all the measured values are well above the classical bound\cite{Kuzmich03} of 2, $g_{12}=100$ being the highest value observed. A value above $\sim$7 is required for violating a Bell inequality as in \cite{deRiedmatten2006} where polarization entanglement is generated with a single node, and a value above $\sim$25 is necessary when two parallel pairs of entangled memories are used for realizing for instance a quantum repeater segment \cite{Chou07}. The reported values are well above these operational benchmarks, which are critical for the practical realization of elaborate quantum networking architectures \cite{Kimble}.  

Given a stored collective excitation, a crucial feature is the retrieval efficiency $q_c$, which measures the ability to efficiently retrieve this excitation as a photon after the read pulse is sent. To access this quantity, we first measure the conditional probability $p_{c}= p(2|1)=(p_{12}/p_1)$ of detecting a guided Field-2 photon after retrieval. This probability is lower than $q_c$ due to the experimental losses in the Field-2 path.  With $\eta_2$ the overall detection efficiency for Field-2 photons, $q_{c}$ and $p_{c}$ are directly related by $q_{c}=p_{c}/\eta_2$. Experimentally we measure $\eta_2=0.14$, a value that includes the filtering system and fibre-connection transmissions as well as the detector efficiency. 

Figure \ref{fig3}b displays the retrieval efficiency $q_{c}$ as a function of $p_{1}$. We can observe three different regimes. In the first one, for a large $p_{1}$, $q_{c}$ increases as $p_{1}$ rises, which is in agreement with the multi-excitation process dependence on the write-field energy. The second region is a plateau in the $q_c$ curve and corresponds to the single excitation regime, as shown later. In this regime, the retrieval efficiency reaches a value of (25$\pm$3)\%. The third region corresponds to very low excitation probability, where the noise background that creates false heralding events becomes predominant and thereby leads to lowered $q_{c}$ values. In the intermediate regime, i.e., $p_1\sim5.10^{-3}$, the experimental heralding rate is 5~kHz in the writing operation phase.  

The points in Fig. \ref{fig3}a and \ref{fig3}b are fitted to a model that assumes the total output field state to be composed of a two-mode squeezed state plus background fields in coherent states corresponding to experimental imperfections (See Methods). These backgrounds can be proportional to the write field intensity, such as the light scattered from the write pulse into the guided mode, or independent of it, such as dark noise and read-beam scattering. The data agree well with this model. 

The retrieval efficiency is not currently limited by the optical depth. Additional measurements show a saturated value for ODs between 20 and 40, and a drop for larger ones. Similar trend was observed in free-space experiments, with efficiencies up to 50\% \cite{Laurat07}. Whereas this trend is originating from the multiple excited levels that lead to an effective source of decoherence, the mechanisms responsible for the difference in efficiency are under investigations and include polarization alignment, which is delicate in our fibred implementation. However, we note that a model of the employed protocol including multi-level atomic configuration has not been realized so far, even in free space to the best of our knowledge. Additionally here, the complex polarization structure of guided modes, i.e., spin-momentum locking of light, and the related possible chiral coupling \cite{Lodahl2015} should be taken into account. This effect will play a major role if all the beams are guided. A full model will be the subject of future work and would provide not only paths for optimization but should also reveal specific phenomena to explore.

\begin{figure}[b!]
\includegraphics[width=0.93\columnwidth]{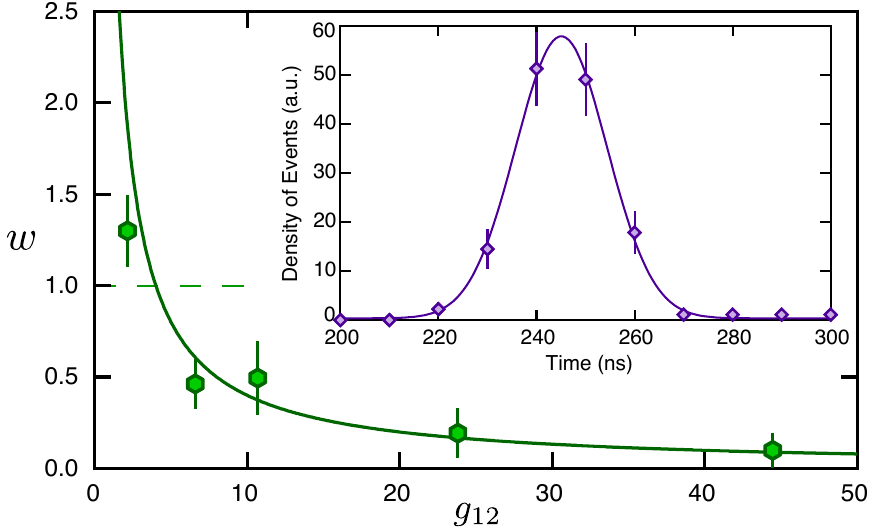}
\caption{\textbf{Characterization of the guided single photon}. The suppression of the two-photon component of the Field 2 obtained from the excitation readout is measured by the antibunching parameter $w$ (green hexagons) and given as a function of $g_{12}$. The lowest $w$ value reaches $0.10 \pm 0.09$ for $g_{12}=45$. The green solid line shows the expected behavior $w\simeq4/g_{12}$ for an ideal two-mode squeezed state. The inset gives the single-photon wave packet measured via the coincidence events (purple diamonds) limited to the 10 ns resolution of our detection system. The solid line represents a Gaussian fit of $1/e$ width 26 ns. The error bars correspond to the propagated Poissonian error of the photon counting probabilities.}
\label{fig4}
\end{figure}

The results presented so far were obtained for a short storage time, i.e., 200~ns. Being stored in a coherent superposition of the ground state levels, the collective excitation can be readout at a later time. To further characterize the system, figure \ref{fig3}c provides the decay of the retrieval probability with the storage time. We obtain a memory lifetime of 3~$\mu$s. This value comes from the residual magnetic field that results in inhomogeneous broadening and decoherence of the stored excitation. The average broadening measured by microwave spectroscopy is on the order of 150~kHz, in agreement with the observed decay. The storage time could be greatly improved by optical pumping of the trapped atoms into a single Zeeman level that will enable to take advantage of the ms-range lifetime of the compensated dipole trap. However, preserving a large optical depth, and thereby an efficient readout, is challenging in this process. We are exploring refined atomic sample preparation and further cooling techniques to combine large optical depth and long storage time.

Finally, to confirm the single character of the heralded excitation and therefore the emission of a single photon into the guided mode after readout, we measure the degree of suppression $w$ of the two-photon component of the retrieved Field 2 compared to a coherent state. For this purpose, a Hanbury Brown-Twiss setup composed of a $50/50$ beam splitter and two single-photon detectors is inserted in the path of the Field-2 photon. Figure \ref{fig4} shows the antibunching value $w$ as a function of the correlation parameter $g_{12}$. The value of $w$ is obtained from single and joint probabilities by the expression $w={(p_{1}p_{1,2a,2b})}/{(p_{1,2a} p_{1,2b})}$ where $p_{1,2a,2b}$ indicates the probability for triple coincidences, and $p_{1,2a}$ ($p_{1,2b}$) the probability for coincidences between detectors $\rm{APD}_{1}$ and $\rm{APD}_{2a}$ ($\rm{APD}_{2b}$). The lowest measured value is  $w=0.10 \pm 0.09<1$ for $g_{12}=45$. This measurement confirms the single-photon character of the retrieved field and therefore that the intermediate plateau observed in Fig. \ref{fig3}b corresponds well to the single-excitation regime. The temporal mode of the guided single photon is given in the inset of Fig. \ref{fig4}. The temporal width is of $26~\rm{ns}$, as expected for a strong square-shaped read pulse with a rise time of about 20 ns. We anticipate that widely tunable wave shapes can be obtained, with duration up to a few microseconds, as observed in a recent free-space implementation, without substantial change in the retrieval efficiency \cite{Farrera2016}.

Our results demonstrate the capability to herald, store, and read out a single collective atomic excitation that is preferentially coupled to a 1D nanoscale waveguide. This capability  brings the emerging paradigm of atomic waveguide-QED  into the quantum regime, and has both technological and fundamental importance. It first enables applications in quantum networks, where this platform can be immediately used as a wired node. Straightforward modifications to our setup would allow write and read beams to be guided, with ultra-low power at the few-photon level. Spin-momentum locking will also lead to interesting investigations in this configuration \cite{Lodahl2015}. Moreover, our realization paves the way to the initialization of preferentially coupled collective excitations in waveguide QED platforms, eventually within a larger space by accumulating excitations and following other heralding schemes, as well as to the exploration of quantum many-body physics in this setting. An important step would be to combine the capability demonstrated here for the still-limited single-atom coupling with ongoing efforts to achieve enhanced interaction in 1D structured waveguides \cite{phc,PNAS}, opening up the prospect of efficient single-pass quantum nonlinear optics protocols. Exciting applications may also arise from the combination of such quantum state engineering of the mesoscopic spin ensemble with potential single-atom addressing in the atomic register.

\vspace{0.3 cm}
\noindent  {\fontfamily{phv}\selectfont 
\normalsize \textbf{Acknowledgements} 
}
\newline
\noindent This work was supported by the European Research Council (Starting Grant HybridNet), the Emergence program from Ville de Paris (Project NanoQIP), the DIM Nano-K from R\'egion Ile-de-France, and the PERSU program from Sorbonne Universit\'e (ANR-11-IDEX-0004-02). N.V.C. and A.S.S. acknowledge the support from the EU (Marie Curie Fellowships Nanofi 659337 and NanoArray 705161) and J.L thanks the Institut Universitaire de France. We also thank D. Maxein, A. Nicolas and O. Morin for their contributions in the early stage of the experiment.

\vspace{0.3 cm}
\noindent  {\fontfamily{phv}\selectfont 
\normalsize \textbf{Author contributions} 
}
\newline
\noindent N.V.C., J.R. and A.C. performed the experiment. B.G. contributed to the preparation of the setup and A.S.S. to the data analysis. All the authors discussed the results and contributed to the writing of the manuscript. J.L. conceived the experiment and supervised the implementation.

\vspace{0.3 cm}
\noindent  {\fontfamily{phv}\selectfont 
\normalsize \textbf{Author Information} 
}
\newline
\noindent The authors declare no competing financial interests. Correspondence and requests for materials should be addressed to J.L. (julien.laurat@sorbonne-universite.fr)

\clearpage
\vspace{0.3 cm}
\noindent  {\fontfamily{phv}\selectfont 
\normalsize \textbf{METHODS} 
}
\newline
\setcounter{figure}{0}
 \renewcommand\figurename{\textbf{Extended Data Figure}}
\small

\noindent In the following, we give a detailed description of the filtering system and present the laser sources and the experimental timing. We then give the characterization of the dipole trap and finally describe the theoretical model used to fit the presented data. \\

\begin{figure}[b!]
\includegraphics[width=0.92\columnwidth]{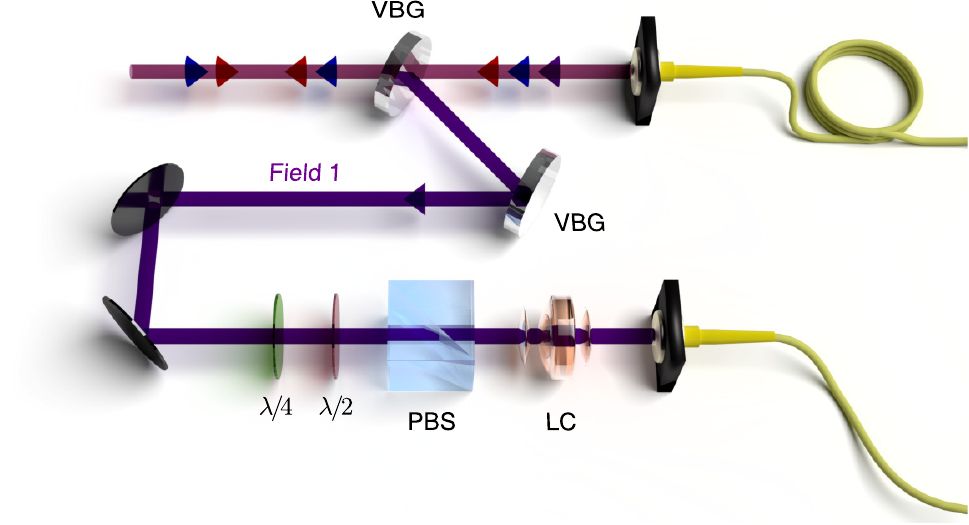}
\caption{\textbf{Filtering system.} The Field-1 beam travels through a filtering system where it is separated from the dipole trapping beams and the write field component coupled to the nanofibre. Two cascaded volume Bragg gratings (VBGs) provide isolation from the dipole beams, while the combination of a polarizing beamsplitter (PBS) and a lens cavity (LC) provides the desired isolation from the write beam. The total transmission of the filtering system is around 40\%, including fibre coupling. A similar filtering system is used for Field 2.}
\label{Extended_fig1}
\end{figure}

\noindent \textbf{Filtering System.} While the dipole trap is based on bright guided lights that are always on, working at the single-photon level for Field 1 and Field 2 is a key capability. Two filtering stages, one on each nanofibre tail, are necessary to reach high isolation. Each filtering system is composed of two cascaded volume Bragg gratings (OptiGrate), a polarizing beam splitter, and a commercial lens-based cavity (Quantaser FPE001B). The cascaded VBGs provide efficient filtering from the dipole beams with an isolation of $120$~dB for a 0.2-nm bandwidth centered around $852$~nm. The lens-based cavity has a transmission of 75 \% with a bandwidth of 80 MHz and a rejection around 40 dB for the write (read) beam in the case of the Field-1 (Field-2) path. The total filtering system transmission reaches 40 \% and includes the coupling into an optical fibre. The filtering system for Field-1 photons is shown in Extended Figure \ref{Extended_fig1} and a similar system is used for Field-2 photons. \\

\begin{figure}[t!]
\includegraphics[width=0.999\columnwidth]{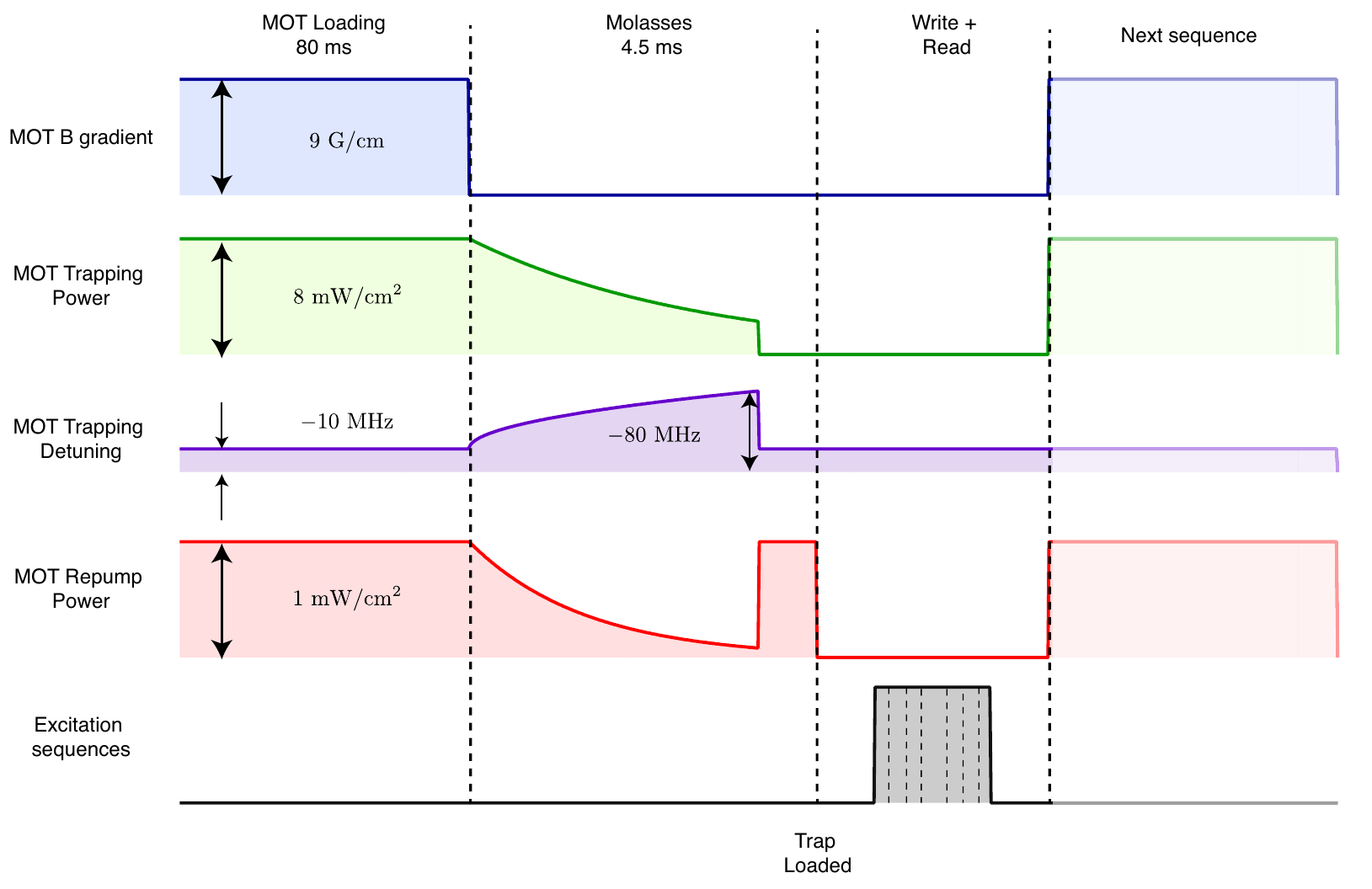}
\caption{\textbf{Timing diagram of the experiment.} After the elongated MOT is loaded, the magnetic field gradient is turned off and the dipole trap is loaded by a transient molasses stage: the MOT-trapping detuning is increased while the MOT-trapping power as well as the MOT-repumping power are decreased. An additional repumping pulse at the end of the molasses stage is sent to the atoms. After this point, the writing and retrieval process starts. The dipole-trapping beams are always on. The full cycle is performed at a repetition rate of 10 Hz.}
\label{Extended_fig2}
\end{figure}

\noindent \textbf{Laser Sources and Experimental Timing.} Different laser sources are used in the experiment. The MOT-trapping and repumping beams are derived from two different laser diodes which are locked on the $\rm{D}_2$ line of $^{133}\rm{Cs}$ (852 nm) via saturated absorption. To obtain the desired power in the trapping beams, a tapered amplifier (Toptica Photonics, BoosTA) is seeded to deliver 100 mW for each beam. The trapping light is red-detuned by 10 MHz from the $|6S_{1/2}, F=4\rangle \rightarrow  |6P_{3/2}, F'=5\rangle$, while the repumping one is resonant to the $|6S_{1/2}, F=3\rangle \rightarrow |6P_{3/2}, F'=4\rangle$ transition. 

A Ti:sapphire laser (MSquared, SolsTiS) is stabilized on a reference cavity and frequency-locked via saturated absorption. From this laser we obtain the write beam, which is red-detuned by 10 MHz from the $|6S_{1/2}, F=4\rangle \rightarrow  |6P_{3/2}, F'=4\rangle$ transition. In addition, this laser gives the probe beam used to characterize the all-fibred dipole trap. A fourth laser diode provides the read beam resonant to the $|6S_{1/2}, F=3\rangle \rightarrow |6P_{3/2}, F'=4\rangle$ transition. This laser is phase-locked to the Ti:sapphire laser at the cesium hyperfine splitting frequency (Vescent Photonics, D2-135). The timing and the detuning of all used laser beams are controlled using acousto-optic modulators (AOMs) in a double-pass configuration.

For the dipole trapping beams, we use three laser diodes (Toptica DLpro). The red-detuned light is at 935 nm, and bounces off a standard dichroic mirror to filter out the residual 852-nm component. Each beam has a power of 0.5 mW. The blue-detuned beams at 686 nm are generated by two different laser diodes because of the targeted relative detuning ($100$~GHz) and each has a power of 4.5 mW. These three lasers are free running and always on.

The timing of the experiment is shown in Extended Figure \ref{Extended_fig2}. The experiment is performed at a repetition rate of 10 Hz. The MOT loading takes 60 ms after which the magnetic field is shut-off. For the next 4.5 ms we apply an optical molasses phase. The detuning of the MOT light is increased from $-10~\rm{MHz}$ to $-80~\rm{MHz}$ with a time constant of 4 ms, and its power decreases as well. The repumping power is also decreased with a 4-ms time constant. A final 0.5 ms-long repumping pulse is sent to ensure the initial state of the atomic ensemble. After this stage, the MOT beams are completely shut-off and the dipole trap is considered loaded. The temperature of the atomic cloud after the molasses stage is measured to be $20~\mu \rm{K}$ and it is obtained via a time-of-flight measurement.

The heralding and retrieval protocol starts 4.5 ms after the shut-off of the magnetic field and lasts a period of $1~\rm{ms}$. For each loading process, we perform between 200 and 1000 protocol repetitions depending on the storage time. Three pairs of bias coils are used to cancel the residual magnetic field. Field-1 and Field-2 photons are detected by avalanche photodiodes (SPCM-AQR-14-FC) and recorded with a FPGA-based digitizer with a 10 ns time resolution. \\

\begin{figure}[t!]
\includegraphics[width=0.99\columnwidth]{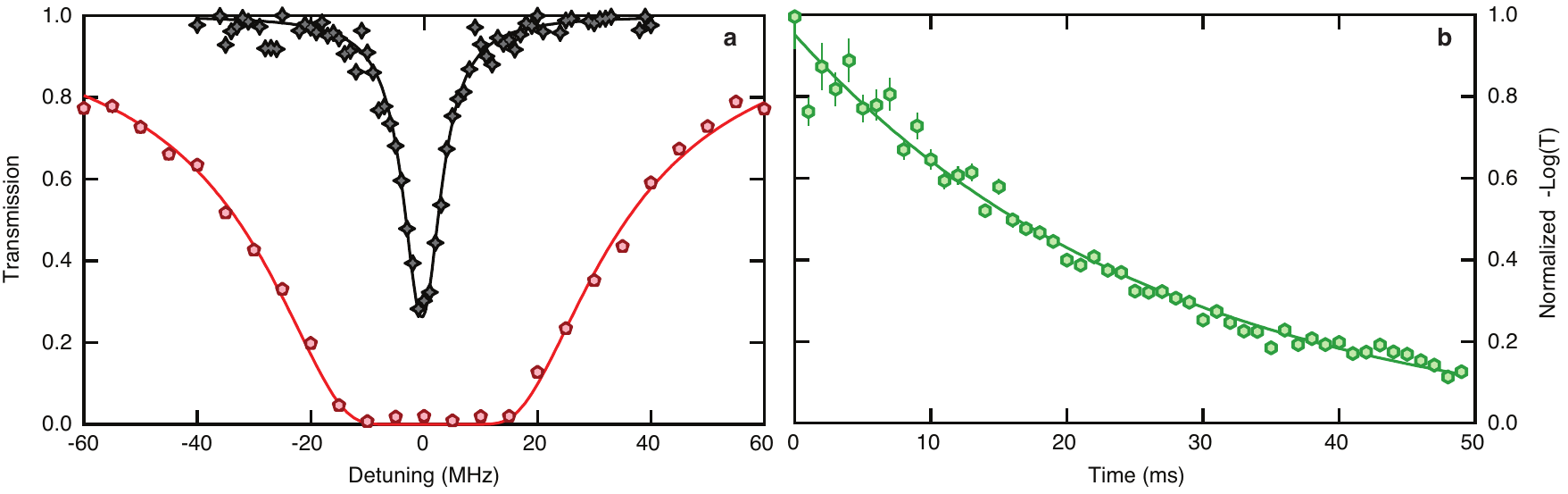}
\caption{\textbf{Dipole trap.} \textbf{a}, Probe transmission as a function of the detuning $\delta$ to resonance for the transition $| 6S_{1/2}, F=4\rangle \rightarrow |6P_{3/2}, F'=5\rangle$. The fits are given by the function $\textrm{Exp}[-\textrm{OD}/(1+(2\delta/\Gamma)^2)]$ with $\textrm{OD}= 97 \pm 2$ (red) and $\textrm{OD}=1.4\pm 0.1$ (black), with $\Gamma/2\pi~=~(5.8 ~\pm~0.2)$~MHz. \textbf{b}, Decay of the absorption after the dipole trap loading. The exponential decay (green solid line) leads to a 25-ms lifetime. The error bars correspond to the propagated Poissonian error of the photon counting probabilities.}
\label{Extended_fig3}
\end{figure}

\noindent \textbf{Dipole Trap Characterization.} The implemented two-color \cite{Balykin} dipole trap follows a compensated, state-insensitive scheme for the $|6S_{1/2}\rangle \rightarrow |6P_{3/2}\rangle$ transition of $^{133}\rm{Cs}$  \cite{nanofiber2}. Both the differential scalar shifts and the vector shifts are strongly suppressed, via the use of the cesium magic wavelengths (686 nm for the blue-detuned beams and 935 nm for the red-detuned ones) and the use of counter-propagating beams, respectively.

Before and after any experimental run, we measure the optical depth of the trapped ensemble by scanning a guided probe pulse on the $|6S_{1/2}, F=4\rangle \rightarrow |6P_{3/2}, F'=5\rangle$ transition, quasi-linearly polarized along the $x$ direction. Extended Figure \ref{Extended_fig3}a shows two different absorption profiles. Optical depths as large as 100 are measured, and can be decreased by reducing the loading time. For low optical depth, we verify the compensation of the dipole trap.

The number of trapped atoms is measured by saturation measurements. For the measurements presented in the main text, with OD$\sim40$ on the $|6S_{1/2}, F=4\rangle \rightarrow |6P_{3/2}, F'=5\rangle$ transition, the total number of atoms is kept to 2000. The lifetime of the trap has been measured to be 25 ms, as reported in Extended Figure \ref{Extended_fig3}b. Due to the tight atom localization in the microscopic traps \cite{nanofiber2,Corzo2016}, the inhomogeneous broadening in the atom-guided mode coupling is limited below 15\%.\\

\noindent \textbf{Theoretical Model.} The experimental results presented in Fig. 3, i.e., $g_{12}$ and $q_c$, are fitted accordingly to a model that assumes the state for Field 1 and Field 2 at the output of the ensemble consists of a two-mode squeezed state,
\begin{equation}
|\Phi_{12}\rangle=\sqrt{1-p}\sum_{n=0}^{\infty}{p^{n/2} |n_{1}n_{2}\rangle},
\end{equation}
plus background fields in coherent states that correspond to the experimental imperfections \cite{Chou04,ChouThesis}. We distinguish two backgrounds $|\nu_{1,2}\rangle$ proportional to the write field power and two incoherent backgrounds $|\nu_{1b,2b}\rangle$ to account for processes that do not depend on this power. The total field at the output of the atomic ensemble is thereby written as:
 \begin{equation}
 |\Psi_{12} \rangle=|\Phi_{12} \rangle |\nu_{1}\rangle |\nu_{2}\rangle |\nu_{1b}\rangle |\nu_{2b}\rangle.
 \end{equation}
Because the read power is fixed, $|\nu_{1,2}\rangle$ account mainly for light scattered from the write field and background fluorescence. We thus assume $|\nu_{1,2}|^2 = \kappa_{1,2}p$. The incoherent backgrounds $|\nu_{1b,2b}\rangle$ account for light scattered from the read beam and detector dark noise.

This simple model enables to obtain the expressions for single and joint probabilities of detecting Field 1 and Field 2. The single probabilities can be written as:
\begin{eqnarray}
p_{1}&=&\alpha_{1}\langle \hat{n}_{1}\rangle =\alpha_{1}\bigg(\frac{p}{1-p}+|\nu_{1}|^2+|\nu_{1b}|^2\bigg)\nonumber\\
p_{2}&=&\alpha_{2}\langle \hat{n}_{2}\rangle =\alpha_{2}\bigg(\frac{p}{1-p}+|\nu_{2}|^2+|\nu_{2b}|^2\bigg)\nonumber
\end{eqnarray}
where $\alpha_{1,2}$ are constants giving the efficiencies of converting the respective field modes into photodetections. These coefficients take therefore into account the detection efficiency and the transmission losses. The joint probability is given by:
\begin{eqnarray}
p_{12}&=&\alpha_{1}\alpha_{2}\langle :\hat{n}_{1}\hat{n}_{2}:\rangle \nonumber\\
&=&\alpha_{1} \alpha_{2} \bigg( \frac{p+p^2}{(1-p)^2} +\frac{p}{1-p}\big(|\nu_{1}|^{2}+|\nu_{1b}|^{2}+|\nu_{2}|^{2}+|\nu_{2b}|^{2}\big) \nonumber\\
&&+\big(  |\nu_{1}|^{2} +|\nu_{1b}|^{2}      )(|\nu_{2}|^{2}+|\nu_{2b}|^{2}\big)  \bigg).\nonumber
\end{eqnarray}
The above probabilities provide the cross-correlation function and the retrieval efficiency as:
\begin{align}
g_{12}=\frac{p_{12}}{p_{1}p_{2}}\qquad \textrm{and} \qquad \eta_2\,q_{c}=\frac{p_{12}}{p_{1}}.
\end{align}

For the solid theoretical curves in Fig. 3, we set the value of $\alpha_1=0.16$ which is our measured experimental transmission for Field 1. This value includes the transmission in the filtering system (0.4), the detection efficiency (0.5), and other transmission losses though fibre connections (0.8). The value of $\alpha_2=0.035$ comes from a straightforward fitting in Fig. 3b, since this parameter corrected for the total transmission $\eta_2$ in the Field-2 path ($\eta_2$=0.14) gives the plateau corresponding to the single-excitation regime. The retrieval efficiency  is $q_c=0.035/0.14=0.25$. The values for $|\nu_{1b,2b}|^2$ are obtained from the measured photodetection probabilities $b_{1,2}$ with MOT off, using the relations $|\nu_{1b,2b}|^2=b_{1,2}/\alpha_{1,2}$. We measure $b_1=5.1\times10^{-5}$ and $b_2=1.6\times10^{-4}$. The remaining parameters are obtained from fitting this model to the experimental data. The fitted values are $\kappa_1=0.034$ and $\kappa_2=1.91$. The value of $\kappa_1$ shows that the ``good" heralding photons exceed the coherent background by a factor of more than 30.


\begin{thebibliography}{10}

\bibitem{nanofiber1} Vetsch, E., Reitz, D., Sagu\'e, G., Schmidt, R., Dawkins, S. T. \& Rauschenbeutel, A. Optical Interface Created by Laser-Cooled Atoms Trapped in the Evanescent Field Surrounding an Optical Nanofiber. \textit{Phys. Rev. Lett.} \textbf{104}, 203603 (2010).
\bibitem{nanofiber2} Goban, A. \textit{et al.} Demonstration of a State-Insensitive, Compensated Nanofiber Trap. \textit{Phys. Rev. Lett.} \textbf{109}, 033603 (2012). 
\bibitem{Lukin2013} Thompson, J. D., Tiecke, T. G., de Leon, N. P., Feist, J., Akimov, A. V., Gullans, M., Zibrov, A. S., Vuleti\'c, V. \& Lukin, M. D. Coupling a Single Trapped Atom to a Nanoscale Optical Cavity. \textit{Science} \textbf{340}, 1202-1205 (2013).
\bibitem{phc} Goban, A. \textit{et al.} Atom-light interactions in photonic crystals. \textit{Nat. Commun.} \textbf{5}, 3808 (2014). 

\bibitem{RMP} Chang, D. E., Douglas, J. S., Gonz\'alez-Tudela, A., Hung, C.-L. \& Kimble, H. J. Quantum matter built from nanoscopic lattices of atoms and photons. \textit{Rev. Mod. Phys.} \textbf{90}, 031002 (2018).

\bibitem{Sile} Nieddu, T., Gokhroo, V. \& Nic Chormaic, S. Optical nanofibres and neutral atoms. \textit{J. Opt.} \textbf{18}, 053001 (2016).
\bibitem{Solano17bis} Solano, P., Grover, J. A., Hoffman, J. E., Ravets, S., Fatemi, F. K., Orozco, L. A. \& Rolston, S. L. Optical Nanofibers: a new platform for quantum optics. \textit{Adv. At. Mol. Opt. Phys.} \textbf{66}, 439-505 (2017).
\bibitem{Duan2001} Duan, L.-M., Lukin, M. D., Cirac, J. I. \& Zoller, P. Long-distance quantum communication with atomic ensembles and linear optics. \textit{Nature} \textbf{414}, 413-418 (2001). 
\bibitem{Sangouard2011} Sangouard, N., Simon, C., de Riedmatten, H. \& Gisin, N. Quantum repeaters based on atomic ensembles and linear optics. \textit{Rev. Mod. Phys.} \textbf{83}, 33 (2011).

\bibitem{Kimble} Kimble, H. J. The quantum internet. \textit{Nature} \textbf{453}, 1023-1030 (2008).
\bibitem{Chang2014} Chang, D.E., Vuleti\'c, V. \& Lukin, M. D. Quantum nonlinear optics - photon by photon. \textit{Nat. Photon.} \textbf{8}, 685-694 (2014).

\bibitem{Sennelart} Somaschi, N. \textit{et al.} Near-optimal single-photon sources in the solid state. \textit{Nat. Photon.} \textbf{10}, 340-345 (2016).


\bibitem{Tiecke2015} Tiecke, T.G., Thompson, J. D., de Leon, N. P., Liu, L. R., Vuleti\'c, V. \& Lukin, M. D. Nanophotonic quantum phase switch with a single atom. \textit{Nature} \textbf{508}, 241-244 (2014).

\bibitem{Lodahl2015} Lodahl, P., Mahmoodian, S. \& Stobbe, S. Interfacing single photons and single quantum dots with photonic nanostructures. \textit{Rev. Mod. Phys.} \textbf{87}, 347 (2015).

\bibitem{Turschmann} T\"urschmann, P. \textit{et al.} Chip-Based All-Optical Control of Single Molecules Coherently Coupled to a Nanoguide. \textit{Nano Lett.} \textbf{17}, 4941-4945 (2017).

\bibitem{vanLoo2013} van Loo, A. F., Fedorov, A., Lalumi\`ere, K., Sanders, B. C., Blais, A. \& Wallraff, A. Photon-mediated interactions between distant artificial atoms. \textit{Science} \textbf{342}, 1494-1496 (2013). 

\bibitem{Schleier} Schleier-Smith, M. Hybridizing Quantum Physics and Engineering. \textit{Phys. Rev. Lett.} \textbf{117}, 100001 (2016).

\bibitem{Pichler2015} Pichler, H., Ramos, T., Daley, A. J. \& Zoller, P. Quantum optics of chiral spin networks. \textit{Phys. Rev. A} \textbf{91}, 042116 (2015).
\bibitem{Tuleda2017} Gonz\'alez-Tudela, A., Paulisch, V., Kimble, H. J. \& Cirac, J. I. Efficient Multiphoton Generation in Waveguide Quantum Electrodynamics. \textit{Phys. Rev. Lett.} \textbf{118}, 213601 (2017).

\bibitem{Asenjo2017} Asenjo-Garcia, A., Moreno-Cardoner, M., Albrecht, A., Kimble, H. J. \& Chang, D. E. Exponential Improvement in Photon Storage Fidelities Using Subradiance and ``Selective Radiance'' in Atomic Arrays. \textit{Phys. Rev. X} \textbf{7}, 031024 (2017).
\bibitem{Douglas2016} Douglas, J. S., Caneva, T. \& Chang, D. E. Photon Molecules in Atomic Gases Trapped Near Photonic Crystal Waveguides. \textit{Phys. Rev. X} \textbf{6}, 031017 (2016).

\bibitem{Douglas2015} Douglas, J. S., Habibian, H., Hung, C.-L., Gorshkov, A. V., Kimble, H. J. \& Chang, D. E. Quantum many-body models with cold atoms coupled to photonic crystals. \textit{Nat. Photon.} \textbf{9}, 326-331 (2015).
\bibitem{Choi2017} Dong, Y., Lee, Y.-S. \& Choi, K. S. Waveguide QED toolboxes for synthetic quantum matter. Preprint at http://arxiv.org/abs/1712.02020.



\bibitem{Gouraud2015} Gouraud, B., Maxein, D., Nicolas, A., Morin, O. \& Laurat, J. Demonstration of a Memory for Tightly Guided Light in an Optical Nanofiber. \textit{Phys. Rev. Lett.} \textbf{114}, 180503 (2015).
\bibitem{Sayrin2015} Sayrin, C., Clausen, C., Albrecht, B., Schneeweiss, P. \& Rauschenbeutel, A. Storage of fiber-guided light in a nanofiber-trapped ensemble of cold atoms. \textit{Optica} \textbf{4}, 353 (2015).
\bibitem{Goban2015} Goban, A., Hung, C.-L., Hood, J. D., Yu, S.-P., Muniz, J. A., Painter, O. \& Kimble, H. J. Superradiance for Atoms Trapped Along a Photonic Crystal Waveguide. \textit{Phys. Rev. Lett.} \textbf{115}, 063601 (2015).
\bibitem{Solano17} Solano, P., Barberis-Blostein, P., Fatemi, F. K., Orozco, L. A. \& Rolston, S. L. Super-radiance reveals infinite-range dipole interactions through a nanofiber. \textit{Nat. Commun.} \textbf{8}, 1857 (2017).
\bibitem{Corzo2016} Corzo, N. V., Gouraud, B., Chandra, A., Goban, A., Sheremet, A. S., Kupriyanov, D. V. \& Laurat, J. Large Bragg Reflection from One-Dimensional Chains of Trapped Atoms Near a Nanoscale Waveguide. \textit{Phys. Rev. Lett.} \textbf{117}, 133603 (2016).
\bibitem{Polzik2016} S\o rensen, H. L., B\'eguin, J.-B., Kluge, K. W., Iakoupov, I., S\o rensen, A. S., M\"uller, J. H., Polzik, E. S. \& Appel, J. Coherent Backscattering of Light Off One-Dimensional Atomic Strings. \textit{Phys. Rev. Lett.} \textbf{117}, 133604 (2016).


\bibitem{Kuzmich03} Kuzmich, A., Bowen, W. P., Boozer, A. D., Boca, A., Chou, C.-W., Duan, L.-M. \& Kimble, H. J. Generation of nonclassical photon pairs for scalable quantum communication with atomic ensembles. \textit{Nature} \textbf{423}, 731-734 (2003).

\bibitem{Laurat07} Laurat, J., de Riedmatten, H., Felinto, D., Chou, C.-W., Schomburg, E. W. \& Kimble, H. J. Efficient retrieval of a single excitation stored in an atomic ensemble. \textit{Opt. Express} \textbf{14}, 6912 (2006). 


\bibitem{deRiedmatten2006} de Riedmatten, H., Laurat, J., Chou, C.-W., Schomburg, E. W., Felinto, D. \& Kimble, H. J. Direct Measurement of Decoherence for Entanglement between a Photon and Stored Atomic Excitation. \textit{Phys. Rev. Lett.} \textbf{97}, 113603 (2006).
\bibitem{Chou07} Chou, C.-W., Laurat, J., Deng, H., Choi, K. S., de Riedmatten, H., Felinto, D. \& Kimble, H. J. Functional Quantum Nodes for Entanglement Distribution over Scalable Quantum Networks. \textit{Science} \textbf{316}, 1316-1320 (2007).

\bibitem{Farrera2016} Farrera, P., Heinze, G., Albrecht, B., Ho, M., Ch\'avez, M., Teo, C., Sangouard, N. \& de Riedmatten, H. Generation of single photons with highly tunable wave shape from a cold atomic ensemble. \textit{Nat. Commun.} \textbf{7}, 13556 (2016).

\bibitem{PNAS} Burgers, A. P., Peng, L. S., Muniz, J. A., McClung, A. C., Martin, M. J. \& Kimble, H. J. Clocked Atom Delivery to a Photonic Crystal Waveguide. \textit{Proc. Natl. Acad. Sci. USA} \textbf{116}, 456-465 (2019).
\setcounter{firstbib}{\value{NAT@ctr}}

\end{thebibliography}

\begin{thebibliography}{9}
\setcounter{NAT@ctr}{\value{firstbib}}
\bibitem{Balykin} Le Kien, F., Balykin, V. I. \& Hakuta, K. Atom trap and waveguide using a two-color evanescent field around a subwavelength-diameter optical fiber. \textit{Phys. Rev. A} \textbf{70}, 063403 (2004).
\bibitem{Chou04} Chou, C.-W., Polyakov, S. V., Kuzmich, A. \& Kimble, H. J. Single-Photon Generation from Stored Excitation in an Atomic Ensemble. \textit{Phys. Rev. Lett.} \textbf{92}, 213601 (2004).
\bibitem{ChouThesis} Chou, C.-W. Ph.D. thesis, California Institute of Technology (2006).

\end{thebibliography}
\end{document}